\documentclass{aastex63}
\usepackage{colordvi}
\usepackage{color}
 \newcommand{\B}[1]{{\color{blue}{#1}}}
 \newcommand{\R}[1]{{\color{red}{#1}}}
 
\begin{document}

\title{Microwave Study of a Solar Circular Ribbon Flare}


\author{Jeongwoo Lee}
\affil{Institute for Space Weather Sciences, New Jersey Institute of Technology, University Heights, Newark, NJ, USA}

\author{Stephen M. White}
\affil{Space Vehicles Directorate, Air Force Research Laboratory, Albuquerque, NM, USA}

\author{Xingyao Chen}
\affil{CAS Key Laboratory of Solar Activity, National Astronomical Observatories of Chinese Academy of Sciences, Beijing 100101, China}

\author{Yao Chen}
\affil{Institute of Space Sciences, Shandong University, Weihai, Shandong 264209, China}

\author{Hao Ning}
\affil{Institute of Space Sciences, Shandong University, Weihai, Shandong 264209, China}

\author{Bo Li}
\affil{Institute of Space Sciences, Shandong University, Weihai, Shandong 264209, China}

\author{Satoshi Masuda}
\affil{Institute for Space-Earth Environmental Research, Nagoya University, Aichi 464-8601, Japan}

\begin{abstract}
A circular ribbon flare SOL2014-12-17T04:51 is studied using the 17/34 GHz maps from the Nobeyama Radioheliograph (NoRH) along with (E)UV and magnetic data from the Solar Dynamics Observatory (SDO). We report the following three findings as important features of the microwave CRF. (1) The first preflare activation comes in the form of a gradual increase of the 17 GHz flux without a counterpart at 34 GHz, which indicates thermal preheating. The first sign of nonthermal activity occurs in the form of stepwise flux increases at both 17 and 34 GHz about 4 min before the impulsive phase. (2) Until the impulsive phase, the microwave emission over the entire active region is in a single polarization state matching the magnetic polarity of the surrounding fields. During and after the impulsive phase, the sign of the 17 GHz polarization state reverses in the core region, which implies a magnetic breakout--type eruption in a fan-spine magnetic structure. (3) The 17 GHz flux around the time of the eruption shows quasi-periodic variations with periods of 1--2 min. The pre-eruption oscillation is more obvious in total intensity at one end of the flare loop, and the post-eruption oscillation, more obvious in the polarized intensity at a region near the inner spine. We interpret this transition as transfer of oscillatory power from kink mode oscillation to torsional Alfv\'en waves propagating along the spine field after the eruption. We argue that these three processes are inter-related and indicate a breakout process in a fan-spine structure.

\end{abstract}

\section{Introduction}
Circular ribbon flares (CRFs) occur in a special magnetic configuration where a central parasitic magnetic field is surrounded by closed ribbons with the opposite magnetic polarity, implying an overlying dome-shaped fan separatrix (Masson et al. 2009, Sun et al. 2013). The implied fan-spine configuration  has motivated solar MHD theorists to challenge the observed CRF phenomenologies 
(Lau \& Finn 1990, Schrijver \& Title 2002, T\"or\"ok et al. 2009, Pontin et al. 2013, Rickard \& Titov 1996, Galsgaard \& Nordlund 1997, Galsgaard et al. 2003, Pontin \& Galsgaard 2007, Pontin et al. 2007, Pariat et al. 2009, 2010; Karpen et al. 2017; Wyper et al. 2016, 2017, 2018). On the other hand, the first observational study of CRFs was made using the TRACE 1600 {\AA} UV continuum images of a confined C8.6 flare (Masson et al. 2009). Later H$\alpha$ blue-wing images obtained from the digitized films of Big Bear Solar Observatory (BBSO) were used to study five CRFs exhibiting jets (Wang \& Liu 2012). Hard X-ray (HXR) observations with the Reuven Ramaty High Energy Solar Spectroscopic Imager (RHESSI) were used to study the high-energy electron content of a CRF (Reid et al. 2012). EUV observations with the Atmospheric Imaging Assembly (AIA) instrument on board the Solar Dynamics Observatory (SDO), in combination with field-line extrapolation, suggested additional ideas such as hyperbolic flux tube reconnection (Masson et al. 2017), spine-fan reconnection (Liu et al. 2019) as well as the late phase extreme-ultraviolet (EUV) phases (Woods et al. 2011) in a non-eruptive CRF (Masson et al. 2017) and hot spine loops and the nature of a late phase in terms of a cooling process (Sun et al. 2013). Eruptions of flux ropes embedded inside the CRFs have also been studied (Liu et al. 2013, 2019). 

It is often said that CRFs form an important class of solar flares because they imply truly three--dimensional (3D) magnetic reconnection. This 3D nature is clear in theory but may be harder to identify in observations.  Since all flares are actually 3D, it is not sufficient simply to image a large--scale 3D structure around the reconnection point.  The so-called standard solar flare model is understood within a two-dimensional (2D) framework because the ribbon motion away from the magnetic polarity inversion line (PIL) can adequately be described by a 2D picture (Kopp \& Pneuman 1976, Priest \& Demoulin 1995, Demoulin et al. 1996). A good example is the famous Bastille day flare that exhibited a visually impressive structure, but the resulting arcade of newly-formed loops can still be explained by 2D reconnection. 
On the other hand,  H$\alpha$ brightness running along the circular ribbon in a CRF indicates that the reconnection involves a structure beyond axial symmetry that is not reducible to 2D physics. Such a structure on the scale of quasiseparatrix layer (QSL) is yet unresolved, and more observational tools are needed in order to address the small-scale physics.

This Letter presents the study of a CRF focusing on microwave emission. The value of microwave observations as diagnostics of CRFs is an open question. 
The most commonly cited microwave diagnostic is that, in the case of gyroresonance emission, the observing frequency divided by the effective harmonic number gives the field strength of the outmost, optically-thick region (Gary \& Hurford 2004, Lee et al. 1993a). 
On the other hand, magnetic reconnection studies require information on magnetic topology rather than field strength. 
A shortcoming of microwave images in this respect is that they do not show morphological details as readily as EUV images, which can play an essential role in tracing field lines carrying significant density.
We expect that a couple of other properties of microwaves will be useful in this problem. One is the sensitivity of  microwave radiation to energetic electrons:  gyrosynchrotron radiation can detect small numbers of nonthermal electrons thanks to the presence of magnetic fields (Rybicki \& Lightman 1979). Another diagnostic is polarization, which is closely related to the coronal magnetic polarity (Zheleznyakov,1970; Lee at al. 1993b; Zheleznyakov et al. 1996; Lee et al. 1998).  These two properties may be utilized as a unique diagnostic tool for exploring magnetic reconnection in the fan--spine structure.

\section{Morphologies at Microwave and (E)UV channels}

The target we select for this study is the SOL2014-12-17T04:51 flare that occurred in NOAA active region (AR) 12242 at heliographic coordinates S20E09. In this event, a striking circular ribbon structure is clearly visible in both EUV and microwave images. This event has already attracted several studies, including a magnetic field analysis (Liu et al. 2019) and studies of quasi-periodic pulsations (QPP) in the range 1.2--2.0 GHz from the Mingantu Spectral Radioheliograph (Chen et al. 2019) and of the thermal structure (Lee et al. 2020). Here we focus on the microwave data for the event, obtained at 1.0--9.4 GHz from the Nobeyama Radiopolarimeter (NoRP) and at 17/34 GHz from the Nobeyama Radioheliograph (NoRH). The set of  imaging observations at both microwave and (E)UV wavelengths offers a rare opportunity for studying microwave properties of a CRF.

\begin{figure}[tbh]  
\includegraphics[scale=1.0]{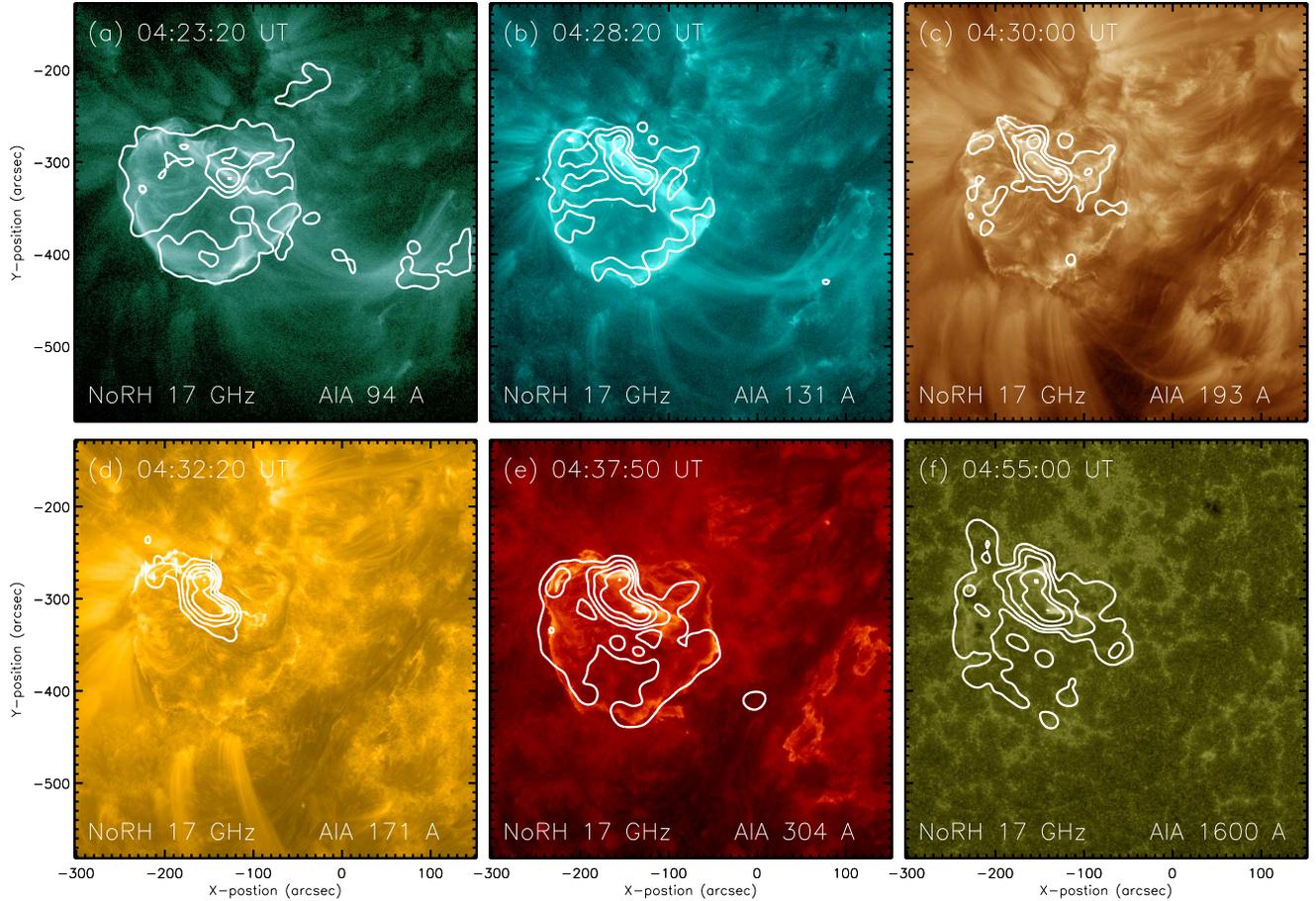}
\caption{EUV and microwave images of the CRF SOL2014-12-17T04:51 in NOAA AR 12242 at six different times: (a)--(c) preflare phase, (d) impulsive phase, and (e)--(f) postflare phase.  Contours in the top panels are at [1.9, 7.0, 26,99]\% of the maximum in each frame, and those in the bottom panels are at [0.5, 1.9, 7.0, 26,99]\% of each maximum. 
Background grayscale images are  AIA images in six different channels at the corresponding times. 
 The animated NoRH 17 GHz and SDO/AIA images include the entire flare event, running from 04:10 to 05:10 UT, with the 17 GHz contour levels at [2, 10, 50, 95]\%  of the maximum in each frame. }
\end{figure}

Figure 1 shows the NoRH 17 GHz maps  as contours at six different times. The background images are all different (E)UV channels at the corresponding times to give an idea of how the CRF appears at different wavelengths. In the accompanying animation, only 94 {\AA} and 131 {\AA} are used as background showing EUV evolution as well as 17 GHz evolution.  In the preflare phase (Fig. 1a--c) the circular structure of the active region is evident.
 (a) The emission is mostly confined within the circular region. Comparison of the 17 GHz and 94 {\AA} images confirms that that microwave emission also outlines the circular-shaped area. The structure inside the circle may be called an ``anemone'' structure.  In the (E)UV channels, the 94 {\AA} image shows a hemispheric structure suggestive of the dome-shaped quasi-separatrix layer (QSL) postulated for CRFs, as does the outer spine structure at the western edge of the frame. 
(b) Close to the flare time, the local region in the north brightens, while the circular ribbon is more obvious in the south of the AR.
In the background EUV images, the outer spine halo structure is best visible at 94 and 131 {\AA} and less apparent in other channels, which means that it is hot and tenuous (Lee et al. 2020).
(c) Near the onset of the impulsive phase, the brightness is more concentrated in an elongated  shape connected to the center of the AR where the strongest magnetic fields are located.
(d) During the flare, the 17 GHz emission is highly concentrated in that region, and the extended source appears to be a flare loop. 
(e) Limited dynamic range in the 17 GHz images makes the southern part of the circular ribbon less prominent in the images when the flare is bright, but it comes back as the flare diminishes. The 304 {\AA} image shows the circular ribbons most clearly. 
(f)  A long decay phase follows during which the EUV source expands and also other areas on the fan surface are visible again at 17 GHz recovering the anemone structure. The 1600 {\AA} image shows the inner and outer flare ribbons where the deposition of the flare energy into the chromosphere is concentrated. They appear to be conjugate footpoints in view of  the loop-like structure in the 17 GHz map (Fig. 1d). 

\begin{figure}[tbh]  
\includegraphics[scale=1.0]{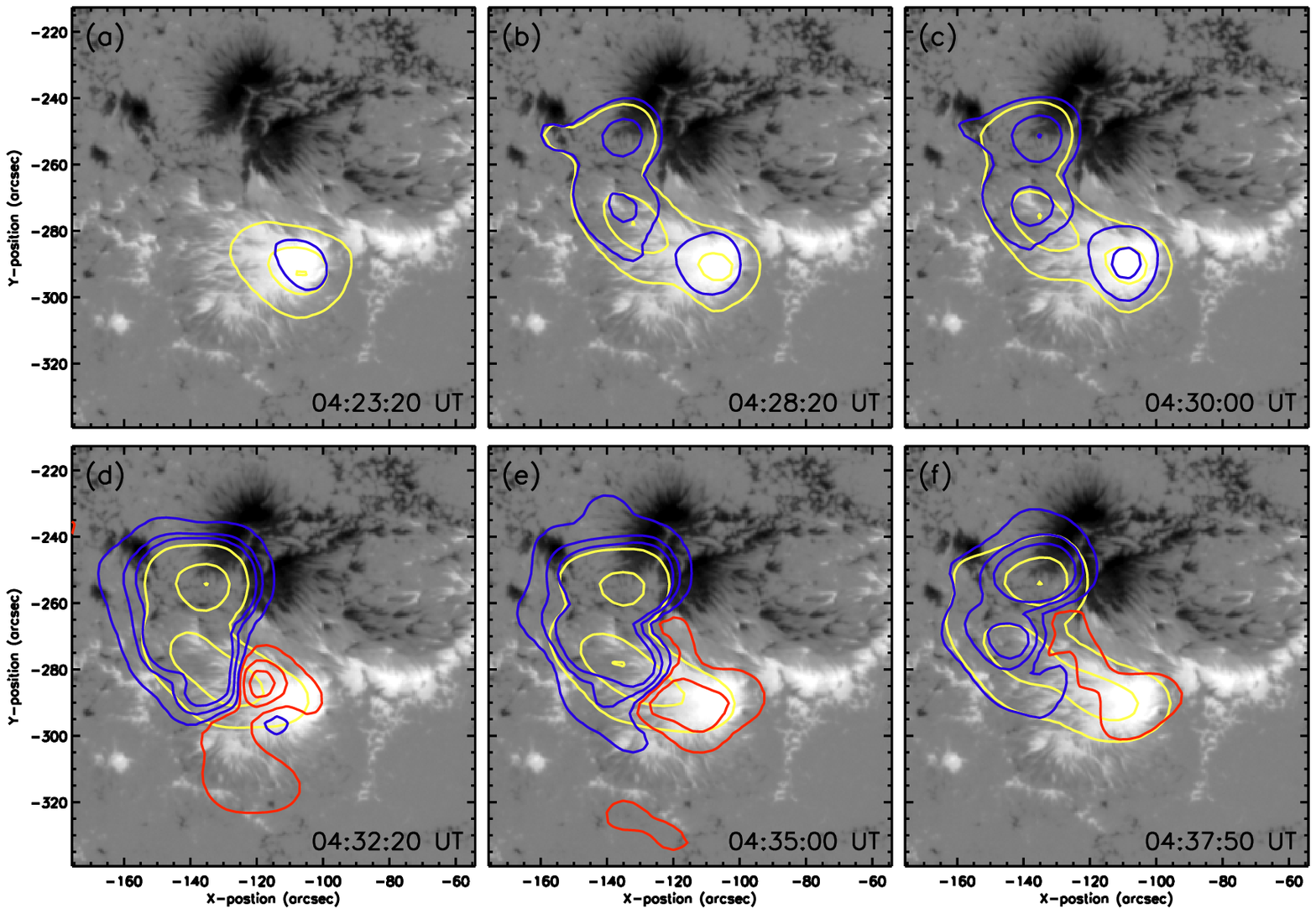}
\caption{Contours of the total and signed polarized intensities at 17 GHz plotted over the HMI line-of-sight magnetograms. The region shown is the main flare site, on the northern edge of the circular ribbon structure evident in Fig. 1. The yellow contours are 17 GHz total intensity plotted at [10, 50, 100]\% of its maximum at each time. The blue (red) contours represent the polarized intensity in LHCP (RHCP) in absolute levels, [10, 50, 100]\% of $\pm$ 2.3 MK.}
\end{figure}

Figure 2 shows contours of the 17 GHz polarized intensity $V=R-L$  at the same times as in Figure 1, except for the last panel.  $V>0$ and $V<0$ are colored red and blue to represent the left-hand circular polarization (LHCP) and the right-hand circular polarization (RHCP), respectively. The total intensity, $I=R+L$, is plotted in yellow contours over the line-of-sight Helioseismic and Magnetic Imager (HMI) magnetograms. The total intensity was initially concentrated over the central sunspot with positive magnetic polarity, and with time expands eastward and also northward to form a loop-like structure in the impulsive phase (Fig. 2a--c). The absence of red contours in Figures 2a--c shows that all sources are LHCP in the preflare phase (top panels), while the region over the positive-polarity sunspot becomes RHCP during the impulsive phase and remains so during decay (Figs. 2d--f). Since the northern sources lie above the negative-polarity region, their natural polarization (corresponding to the extraordinary mode) is expected to be LHCP. However, the central sources over the positive magnetic polarity sunspot should be RHCP (Ratcliffe 1959; Zhelznyakov 1970; Melrose 1975, 1985; Dulk 1985). We thus regard the initial LHCP over this region to be reversed from its nominal polarization, RHCP, in the preflare phase.  This is a new phenomenon, perhaps unique to CRFs, and is likely to be associated with a drastic change in the fan-spine structure, which we discuss further below.

\begin{figure}[tbh]  
\includegraphics[scale=1.0]{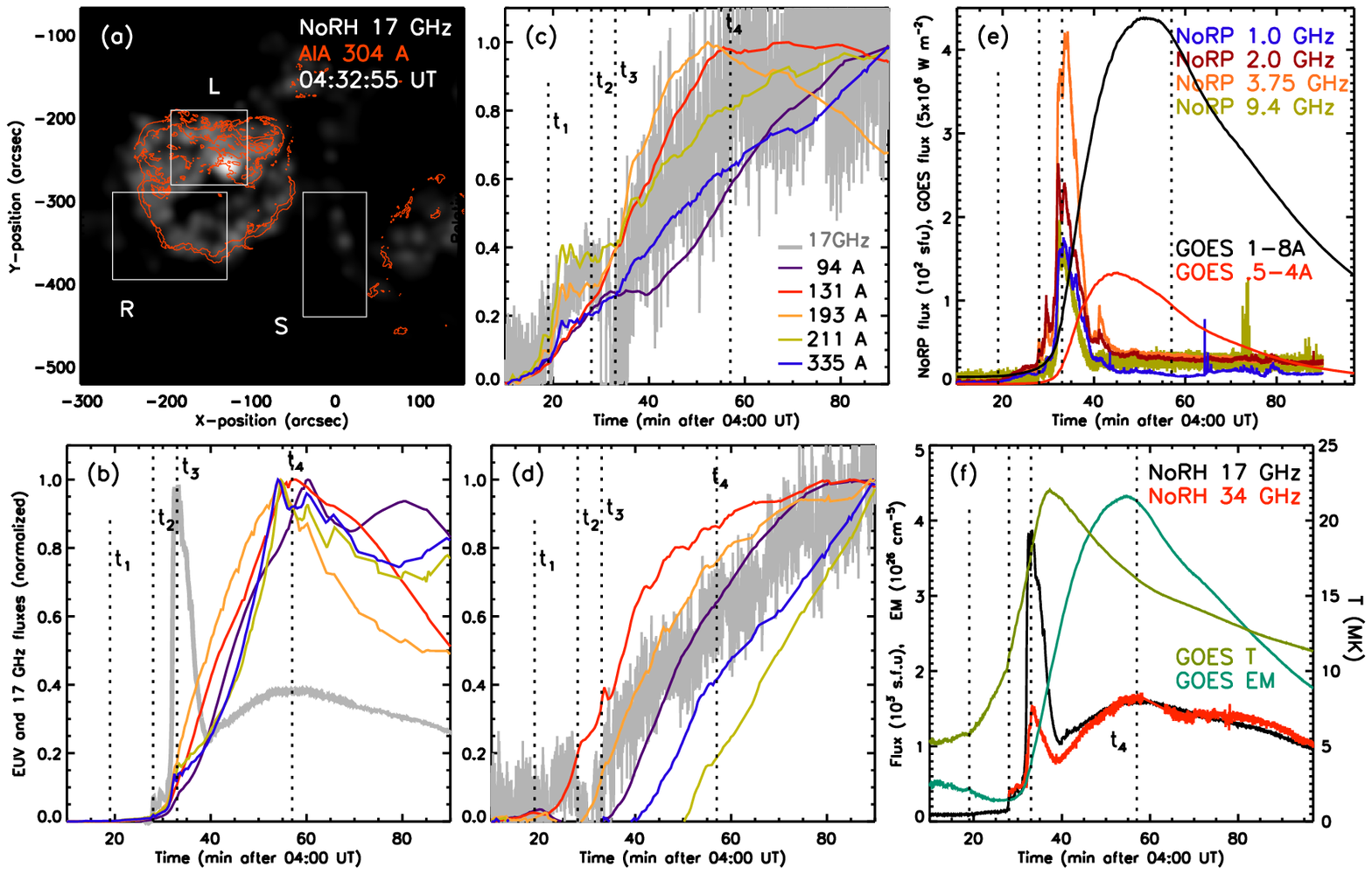}
\caption{Local EUV and 17 GHz fluxes as a function of time: (a) A 17 GHz image (greyscale) showing three regions of interest: the flare loop region ($L$), southern section of the circular ribbon ($R$), and the outer spine region ($S$). The red contours are the  AIA 304 {\AA} intensity at the 10\% level of its maximum. (b)--(d) show the normalized EUV fluxes and the NoRH 17 GHz fluxes versus time, computed for the three regions, $L$, $R$, and $S$, respectively.  $t_1$ and $t_2$ mark the times of stepwise flux variations in L, and $t_3$ is the time of flare maximum at 17 GHz. 
(e) GOES soft X-ray lightcurves are compared with the NoRP (1.0-9.4 GHz) microwave flux time profiles. 
(f) The NoRH 17 and 34 GHz fluxes agree with each other on and after $t_4$ (04:57 UT).  Time profiles of the GOES temperature and EM are also shown.  
Note that the 17/34 GHz fluxes, the EUV fluxes from L, and  GOES EM all simultaneously reach their local maxima at $t_4$.  }
\end{figure}

\section{Correlations with EUV and Soft X-rays}

In Figure 3, we plot the time profiles of microwave fluxes  from NoRP and NoRH along with the AIA EUV and GOES soft X-ray lightcurves. 
Figure 3$a$ shows the three local regions selected for investigation: the flare loop ($L$), the southern section of the circular ribbon ($R$), and the outer spine ($S$). The grayscale image is a 17 GHz map in logarithmic scale down to the 1\% level of the maximum intensity, and the contours are the 304 {\AA} intensity also down to 1\% of its maximum intensity. Note that both of them show that the edge of the enhanced microwave emission coincides with the circular ribbons as represented by the 304 {\AA} intensity. Although microwave maps do not show the circular ribbon itself, it can be inferred from the boundary of enhanced emission from the hotter plasma inside the fan.
Thus any locally enhanced features at 17 GHz are superimposed on a faint background circular disk produced by hot and dense plasmas inside the fan surface. 

Figure 3$b$ shows the fluxes from $L$, which dominate over those from the other regions. The flux time profiles at 17 GHz and five EUV lines at 94 {\AA}, 131 {\AA}, 193 {\AA}, 211 {\AA}, and 335 {\AA} are all normalized to unity, since we are mainly interested in relative timing.  The prominent feature in $L$ is a strong impulsive 17 GHz peak followed by gradually-increasing EUV fluxes and a secondary peak in the 17 GHz flux at the time of the EUV maxima. The impulsive 17 GHz peak is therefore attributed to nonthermal gyrosynchrotron emission by accelerated electrons. The coincidence of the secondary 17 GHz peak at 04:57 UT ($t_4$) with the EUV maxima supports the idea that the gradual 17 GHz flux around $t_4$ is thermal. 
Figures 3$c,d$ shows fluxes from $R$ and $S$, respectively.  The 17 GHz fluxes of these regions are weaker so that the signal to noise ratio (SNR) is lower than in $L$. Some of the EUV fluxes from $R$ show a stepwise flux enhancement at $t_1$ (04:19 UT), signaling the circular ribbon activation. Those fluxes remain enhanced for about 9 min and start to rise again at 04:28 UT ($t_2$). The flux enhancement during $t_1\leq t\leq t_2$ is mostly in the relatively low-temperature passbands, 211 {\AA}, 335 {\AA}, and 304 {\AA} as well as 193 {\AA}, implying thermal emission at standard coronal temperatures. In $S$ (Figure 3$d$), all EUV  fluxes rise at nearly constant rates  with the 131 {\AA} flux leading and the 211 {\AA} flux rising last. The different AIA channels typically have responses peaking at several temperatures, so interpreting the order of the different AIA channels is not straightforward: e.g., 131 \AA\ has both Fe VIII and Fe XXI, so the early rise in 131 \AA\ could be either hot or cold material: for a flare, it makes more sense to assume a dominant hot component. It might be that the rising magnetic field lines carry hotter plasma and the subsequently cooler plasma follows from behind.  The 17 GHz flux is consistent with this trend of EUV fluxes. Thus the results in Figure 3$b$--$d$ indicate that the local microwave fluxes match the behavior of their EUV counterparts when thermal emission dominates. The impulsive nonthermal 17 GHz emission is a tracer of chromospheric heating by precipitating nonthermal electrons, and the consequent rise in EUV emissions, increasing at the onset of nonthermal radiation and continuing to rise afterwards, is consistent with the well known Neupert effect (Neupert, 1968).

Figure 3$e,f$ show the evolution of multiple microwave frequencies and soft X-rays.
In Figure 3$e$, the lower-frequency (non-imaging) NoRP fluxes show impulsive peaks concentrated around $t_3$ as in the NoRH 17 GHz flux from $L$, but the lower frequencies do not show the second peak at $t_4$. A plausible interpretation is that at these frequencies the radio emission is optically thick and therefore the flux represents the temporal evolution of effective temperature, whereas the optically thin emission at $\geq$17 GHz traces the total emission measure of thermal electrons present in the source.
In line with this interpretation, the GOES soft X-ray fluxes start to rise at about $t_2$ and show the fastest variation, indicating the primary energy release, at $t_3$ (Neupert 1968). 
It is also notable that the GOES SXR peaks occur around $t_4$ where the second maximum of microwave flux is observed. Microwave emission with this type of time profile consisting of an impulsive burst followed by a gradual burst at 17 and 34 GHz has been suggested to be a signature for compound flares (Ning et al. 2018).
Figure 3$f$ shows the 17 and 34 GHz fluxes from $L$ computed from the NoRH maps and the temperature and emission measure (EM) calculated from the GOES soft X-rays. The GOES EM peaks around $t_4$, while temperature reaches its maximum much earlier and decreases monotonically through $t_4$. 
The 34 GHz flux is lower than the 17 GHz flux all the way until around $t_4$, after which they are very similar. A flat spectrum is expected when the radiation is dominated by optically-thin thermal free-free emission (e.g., Dulk 1985), and the time profile is mostly dominated by the EM variation. 
The above finding of the EUV line fluxes simultaneously reaching their maxima at $t_4$ (Fig. 3b) is also in line with the conclusion that the microwave emission during the second maximum at $t_4$ is mainly a thermal free-free emission.

Despite the lower SNR of the 17 GHz intensity in $R$ and $S$, this comparison sheds some light on the relationship between the local microwave emissions and the corresponding EUV emissions, which represent different thermal components. It appears that the NoRH 17 GHz fluxes may show spatio-temporal variations similar to those of the AIA EUV fluxes when the 17 GHz fluxes are dominated by thermal emission (e.g., at $t_1$ and $t_4$). When nonthermal emission is dominant ($t_2$ and $t_3$), the 17 GHz fluxes show impulsive behaviors but the EUV fluxes increase only gradually, reaching their maxima much later. 

\begin{figure}[tbh]  
\includegraphics[scale=1.0]{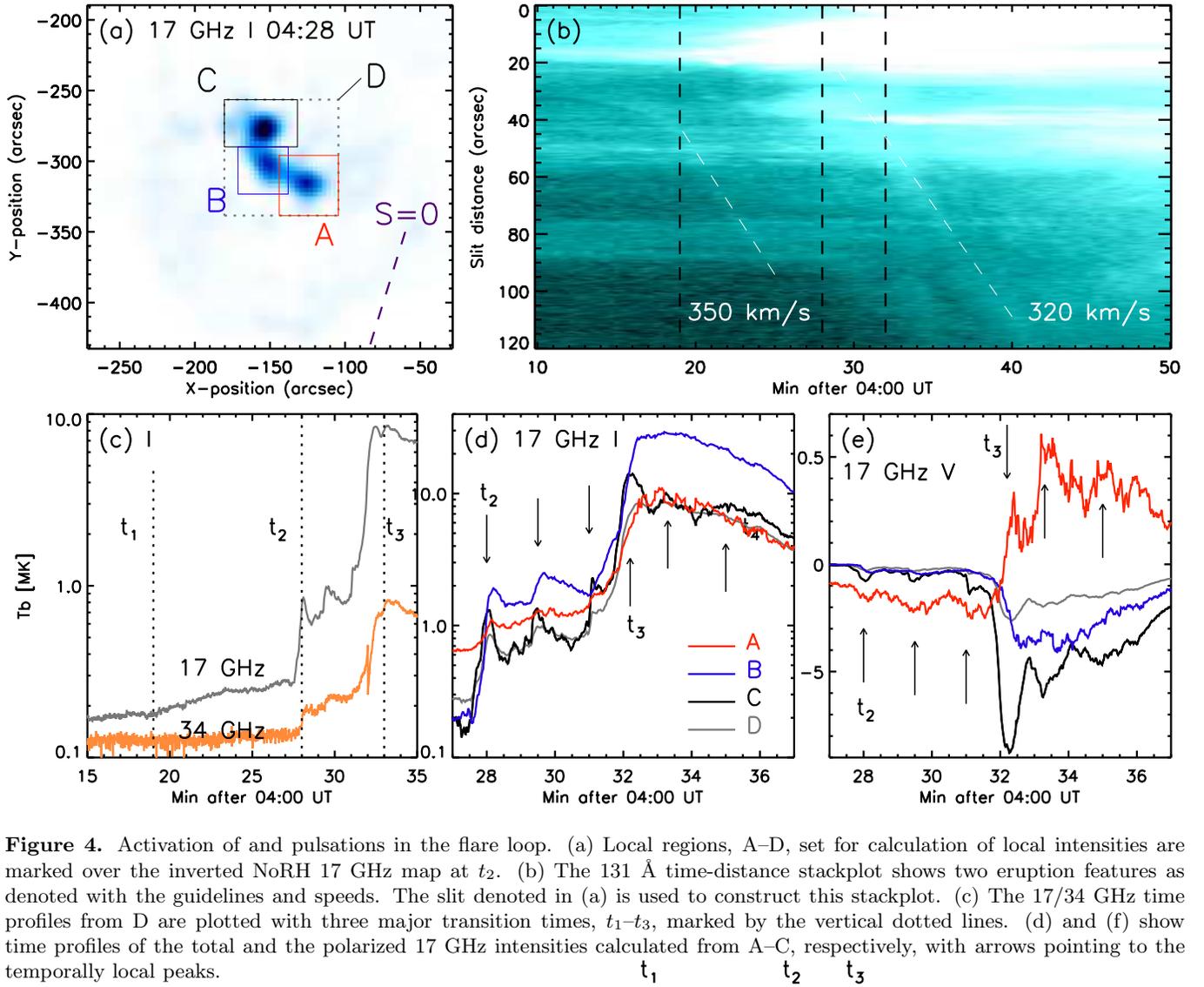}
\caption{Activation of and pulsations in the flare loop. 
(a) Local regions, A--D, set for calculation of local intensities  are marked over the inverted NoRH 17 GHz map at $t_2$.
(b)  The 131 {\AA} time-distance stackplot shows two eruption features as denoted with the guidelines and speeds. The slit denoted in (a) is used to construct this stackplot.
(c) The 17/34 GHz time profiles from D are plotted with three major transition times, $t_1$--$t_3$, marked by the vertical dotted lines.  (d) and (f) show time profiles of the total and the polarized 17 GHz intensities calculated from A--C, respectively, with arrows pointing to the temporally local peaks. }
\end{figure}

\section{Fine Structures in the Flare Loop Activities}

We investigate fine structures in the localized time variations of the microwave bursts by focusing on the strongest emission region $L$. As shown in Figure 4$a$, we set four subregions marked on the inverted 17 GHz intensity map at $t_2$.  A and C are presumably the loop footpoints conjugate to each other, and B is likely to be the looptop location. D includes all the three regions and thus the flare loop. To calculate local flux from each region, we add up all brightness temperatures over the region, and divide it by the number of pixels within the region. The quantities shown in Figure 4 therefore correspond to spatially--averaged local brightness temperatures ($T_b$).  We also mark a slit (dashed purple line) for constructing the time-distance stackplot of the 131 {\AA} intensity displayed in Figure 4$b$. The slit distance starts from the tip denoted as $s=0$ and increases southward to extend over the total distance of 120$''$. In this time-distance stackplot, two eruption features are noticeable, although faint. Their speeds are estimated as $\sim$300 km s$^{-1}$ or higher as denoted by the guidelines. The first feature is likely to have started at $t_1$, suggesting that this eruption may be related to energy release responsible for the initial ribbon activation. Start time of the second eruption feature could be either $t_2$ or $t_3$, which we can hardly discern because of the bright features on the stackplot. In any case, we note that there are indeed eruption-like EUV features corresponding to the CRF activation times, $t_{1,2}$, or the most intense energy release time, $t_3$ (cf. Liu et al. 2019).

Figure 4$c$ shows the 17 GHz and 34 GHz $T_b$ calculated over D. The 17 GHz $T_b$ starts to rise at $t_1$, which we counted as the first activation time based on the EUV lightcurves (Figure 3). Note however that at this time there is no corresponding increase in the 34 GHz $T_b$. The second activation occurs only at $t_2$ in the form of an impulsive increase of both 17 GHz and 34 GHz $T_b$. 
The two activations at $t_1$ and $t_2$ may indicate different levels of energy release, with the latter being more intense. During the flare,  both 17 GHz and 34 GHz $T_b$ impulsively rise at $t_3$. 
Shown in the rest two panels are $T_b$ for the total (Fig. 4$d$) and polarized intensities (Fig. 4$e$)  measured from the subregions, A--C.
The multiple peaks on these time profiles, as marked by the arrows, yield an impression of oscillations superimposed on the flare lightcurve, very similar to the phenomenon called quasi-periodic pulsations (QPPs) as thoroughly studied by Chen et al. (2019) for this event. 
The periodicity is not particularly clear and we would not claim the multiple peaks as QPPs, but instead call them quasi-oscillations.
The quasi-oscillation in total intensity, $I$ (Fig. 4$d$) is more obvious in source C, and less obvious in A. Namely, the farther from the inner spine, the more clearly the quasi-oscillation is visible.
The polarized intensity, $V$ (Fig. 4$e$) show a similar behavior with those of $I$ but with a few differences. The quasi-oscillation of $V$ appear not only in the preflare phase but continue and are stronger during the impulsive phase. Spatially, the quasi-oscillation of $V$ are more obvious in A,  whereas the quasi-oscillation of $I$ are stronger in C.

We were able to count up to five peaks in the lightcurves and measure the time separations between the adjacent peaks. The mean and standard deviation come out as  1.3$\pm$0.2 min for the total intensity in C and 1.7$\pm$0.4 min for the polarized intensity in A. Other combinations exhibit  less number of multiple peaks and their periodicity was not studied. These quasi-oscillations start at $t_2$ with a tendency that the former is more obvious in the start and  the latter lasts longer.
As a comparison, Chen et al. (2019) reported 2 min QPPs in the frequency range 1.2--2.0 GHz around the flaring region during the impulsive phase, 3 min EUV QPPs along the circular ribbon during the preflare phase, and  2 min UV QPPs near the center of the active region from the preflare phase to the impulsive phase (04:00 to 04:45 UT). 
In terms of the spatial location,  the 17 GHz oscillation power residing in the flare loop is an almost identical result with those of the QPP sources at 2 GHz and UV channels  (Chen et al. 2019), except that we used a higher resolution to resolve the loop structure.  The EUV QPP are found in the circular ribbons of the AR, different from other QPP sources.  However, we must note that the oscillations in the 17 GHz and UV radiations are found  in the flare loop region, since those emissions are mostly concentrated there. By contrast,  the EUV emission in the flare loop region is so strongly saturated that the EUV QPP could not be detected there anyway. Therefore we presume that the oscillatory phenomenon is everywhere, but the different locations of QPP at EUV channels and 17 GHz may simply be a matter of which region is more favorable for detecting subtle variations of the radiation.

\begin{figure}[tbh]  
\includegraphics[scale=1.0]{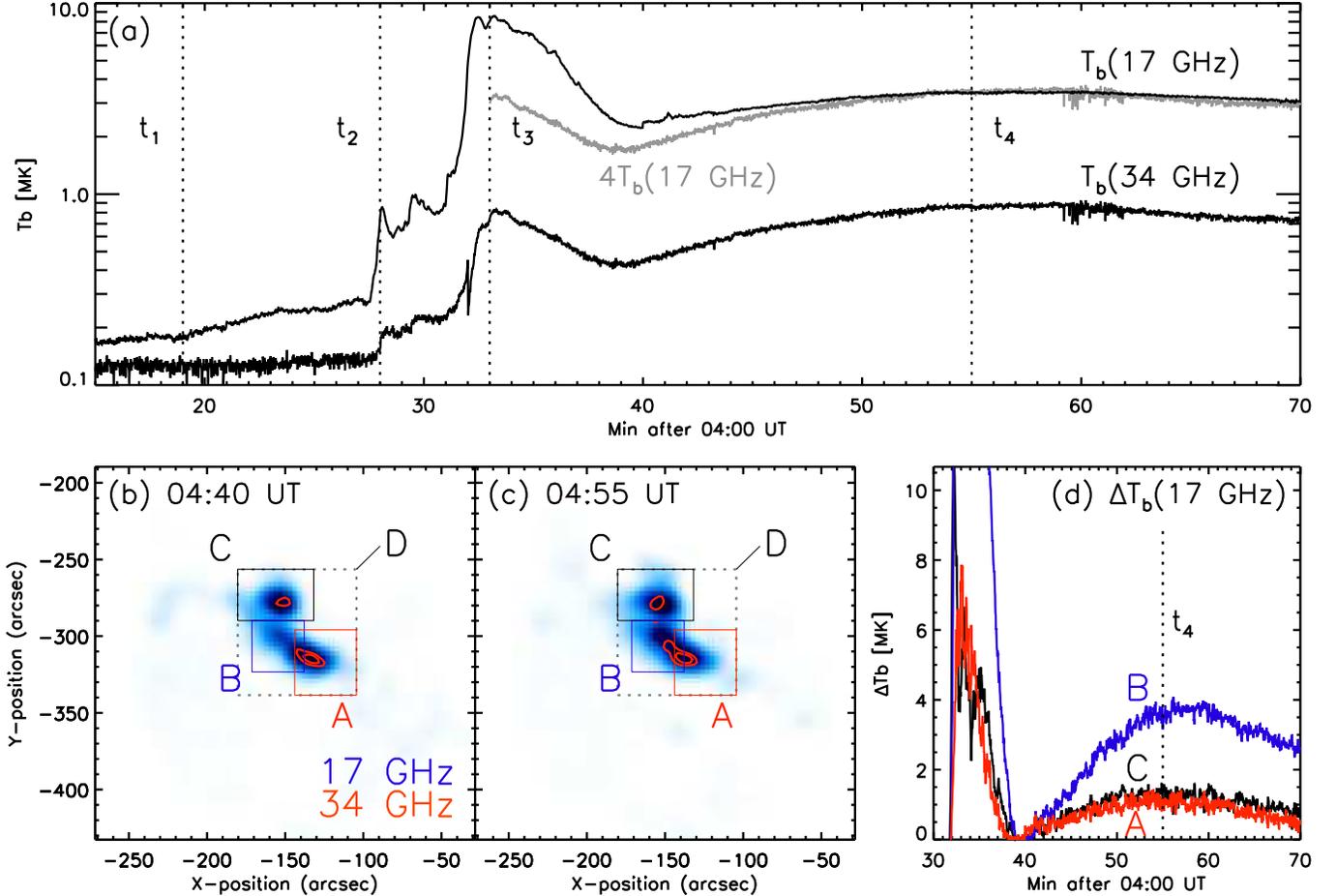}
\caption{
The late phase. (a) The time profiles of $T_b$ at 17/34 GHz from D are plotted along with the four major transition times, $t_1$--$t_4$, marked by the vertical dotted lines.  (b) NoRH 17 GHz map at the start time of the second rise is shown as inverted grayscale image and the 34 GHz map as red contours in the levels of [50, 75]\% of its maximum $T_b$. (c) Same as (b) for the second maximum time.
(d) Time profiles of the relative $T_b$ increase in the three local regions show that the largest $\triangle T_b$ occurred in B.}
\end{figure}

\section{Late-Phase Microwave Activity}
We finally explore the late-phase. Figure 5(a) shows time profiles of $T_b$ at the 17 GHz and 34 GHz averaged over region D. Four key transition times are marked again: $t_1$ for the thermal activation, $t_2$ for the nonthermal activation, and $t_3$ for the maximum number of nonthermal electrons. The late phase activity of this event is then characterized by the gradual rise and fall of $T_b$ around $t_4$, the time of the second maximum $T_b$.  Note that the four-fold $T_b$ at 34 GHz (gray colored curve) tends to agree to the $T_b$ at 17 GHz after $\sim$04:45 UT, which indicates the dominance of optically-thin free-free emission in the late phase.

Figure 5(b--c) show the 17 GHz (grayscale) and 34 GHz (red contours) maps at two different times, both of which generally coincide with each other. These maps at two different times show that the flaring source expands with time toward the late phase. Such an expansion, which the EUV images also show, can be regarded as a CRF version of the expanding flare arcade.
Figure 5(d) shows the relative brightness temperature enhancement ($\triangle T_b$) in the three local regions in reference to the minimum brightness time between $t_3$ and $t_4$.   Since the dominant radiation mechanism at time period is optically-thin free-free emission, $\triangle T_b$ is a measure. The result that $\triangle T_b$ in the looptop (B) is larger than in the footpoints (A or C) indicates high density accumulation in the looptop at $t_4$. This result is consistent with the finding that the  local EUV fluxes in the flare region reach their maxima at $t_4$ (Fig. 2b). 

Post-flare microwave emission from the top of a flaring loop has been detected in many events, and interpreted to be  due to several reasons: trapping of nonthermal electrons in flare loops (Reznikova et al. 2009), enhancement of plasma flows along supra-arcade structures (Kim et al. 2014), and strong heating near the X-point (Chen et al. 2016, 2017). However, none of these may apply to the present event, because this second flux enhancement is distinctively well separated from the impulsive peak ($t_4-t_3\approx 25$ min). Such a property can be more appropriately considered within the context of the EUV late phase activity (Woods et al. 2011; Hock et al. 2012).  Phenomenology, the secondary microwave flux enhancements coincident with the GOES soft X-ray peaks was simply regarded as late phase  thermal activity in compound flares (Lee et al. 2017, Ning et al. 2018).

\section{Discussion}

We have studied the circular ribbon flare, SOL2014-12-17T04:51, mainly using the 17/34 GHz NoRH maps to find new properties inherent to microwave radiation. They are: 
(1) two activation times detected in the form of flux increases at 17 GHz and 34 GHz,
(2) 17 GHz polarization sign reversal at the flare maximum time,
(3) 17 GHz QPP of  $I$ in the preflare phase and QPP of  $V$ in the flare concentrated in different locations, and
(4) the second maximum of 17/34 GHz fluxes is due to a density increase in the flare loop, which has implications for the nature of the late EUV phase study (cf. Lee et al. 2020). We mainly discuss the first three results in relation to the eruption.

\subsection{Preflare Activations: thermal and nonthermal}

Impulsive microwave bursts during the flare can be regarded as nonthermal gyrosynchrotron radiation. But the preflare and the postflare activities may be due to other mechanisms, which include free-free emission and gyroresonant radiation  (Dulk 1985). 
The earliest ribbon activation at $t_1$ in this event comes in the form of a gradual rise of the 17 GHz flux without being accompanied by the 34 GHz flux. Such a distinct response at two separate frequencies cannot be explained by either the thermal free-free or the nonthermal gyrosynchrotron mechanisms, because the spectra of these radiations are broadband (Zheleznyakov 1970). The very discrete frequency dependence is a characteristic of the thermal gyroresonance mechanism (Zheleznyakov 1962, Zheleznyakov \& Zlotnik 1964, 1988). It occurs because gyroresonance opacity is limited up to a few low harmonics ($n=f/f_B=1,2,3,4$ where $f_B=2.8 B_g$ MHz is the gyrofrequency of electrons and $B_g$ is field strength in gauss) above which the opacity drops significantly. Therefore, for a given frequency (in the present case either 17 or 34 GHz) there is a minimum magnetic field required to make the local opacity significant, given by $B_g\geq f$[GHz]$/2.8n$ (see, for further explanations, Gary \& Hurford 2004). The effective harmonic number, $n$, is determined by temperature and density. For a typical coronal temperature ($\sim$1 MK) and density ($10^9$ cm$^{-3}$), $n=3$ is the highest harmonic that has significant opacity. On very vigorous active regions with higher temperatures, $n=4$ may also have significant opacity (White et al. 1992; Lee et al. 1993ab; White \& Kundu 1997). 
For 17 GHz, this means that the local coronal magnetic field strength should be above 1350 gauss, and for 34 GHz, 2700 G is required. Per field strengths from the HMI magnetogram, the former field strength is possibly available in the inner spine, but the latter is not. 
Note, however, that thermal gyroresonance opacity even at 34 GHz was reported for a record-breaking strong coronal magnetic field (Anfinogentov et al. 2019).
This property of gyroresonance opacity explains how the activation at $t_1$ can be seen at 17 GHz, but not at 34 GHz. On the other hand, the second activation at $t_2$ occurs simultaneously at 17 GHz and 34 GHz. This can be explained by nonthermal gyrosynchrotron emission, which is emitted over a wide range of higher harmonics (Zheleznyakov 1970). 
The thermal and nonthermal nature of the two activations is also consistent with the temporal behaviors in that the 17 GHz flux increases gradually at $t_1\leq t \leq t_2$ and impulsively at $t_2$.

\subsection{Rapid Change of Microwave Polarization}

The 17 GHz polarization reversal during this CRF can be a yet unknown feature inherent to the fan--spine structure, where magnetic polarity around the null point varies so rapidly as to affect the propagation of microwave polarization.  A way to possibly explain this polarization change is to view it as a mode-coupling phenomenon, the process by which the rays reverse their original sense of polarization while passing through a quasi-transverse field region along the line of sight from the radiation source to the observer, depending on the degree of mode coupling there (Cohen 1960, Zheleznyakov 1970, Melrose 1975, White et al. 1992). This is an attractive scenario for a fan--spine structure, because the fan surface may well act as a  quasi-transverse layer for the rays emitted underneath.  
To think about an ideal fan-spine structure with a flux rope inside, in this configuration, the magnetic fields above the fan surface are all in the negative magnetic polarity, and the rays emitted from either magnetic polarity underneath will be observed as LHCP everywhere.  Therefore, the LHCP observed everywhere before the flare can simply be due to the fan-spine structure, without any strong mode-coupling phenomenon. On the other hand, if a magnetic flux rope rises to reconnect with the overlying fan field, the fan surface may partially open up to let the flux rope erupt out. 
Such a change of magnetic field structure can explain the instant reversal of the 17 GHz polarization at $t_3$ more naturally.  
The reconnection between the magnetic fields inside and outside of the fan will occur across a current sheet, the so-called breakout current sheet (BCS), and the newly open field lines amount to the lower part of the rising and expanding BCS (see, e.g., Lynch et al. 2016, Karpen et al. 2017). A sustained BCS over the active region might affect the microwave polarization, as mode coupling across a current sheet is still a debatable issue (Zheleznyakov et al. 1996; Lee et al. 1998; Lee 2007). We here offer only the simplest interpretation, according to which the change from LHCP to RHCP of the 17 GHz emission over the inner ribbon is not just a signature for any magnetic field perturbation, but may indicate a specific form of a breakout eruption out of the closed fan structure. The implied magnetic field reconfiguration is in line with the recently reported decay of the coronal magnetic field at the flare site by Fleishman et al. (2020).

\subsection{Trigger of the Eruption}

The start time of the oscillatory behaviors at microwavelengths may yield an implication on the trigger of the eruption. The oscillation itself could start for many reasons. It could have occurred as a dynamic response of the fan-spine system to the eruption, or due to intermittently repeating episodes of flare energy release. A combination of these two is also possible in that the MHD oscillation of the loop can lead to periodically triggered reconnection (Nakariakov \& Melnikov 2009).  In all of these cases, the start time must be $t_3$. However, it is $t_2$ that the 17/34 GHz oscillations started, at which the first eruption signature in the 131 {\AA} channel also started  (Fig. 4).  Other important transitions at $t_2$  are also reported by  independent studies: the start time of the 2 GHz QPP (Chen et al. 2019) and that of the eruption signature in 94 {\AA} (Liu et al. 2019). Their quasi-periods lie in the range of 1.3--4.0 min and are in proportion to the length of loops. It is thus likely that an external driver was applied to this fan-spine structure at $t_2$ and all closed field lines within the dome underwent the kink oscillations (Aschwanden et al. 2002, Zhang et al. 2020), which caused the null point to deform itself into current sheet, and in about 10 min, the eruption broke out (see Lee et al. 2020).
These oscillations started before the eruption and continued after it, which suggests that the erupted field lines serve as a conduit for the waves propagating along the spine.  Among many numerical simulations for fan-spine reconnection (Karpen et al. 2012, 2017; Pariat et al. 2009, 2010, 2015, 2016;  Wyper et al. 2016, 2017, 2018), the latest works (Wyper et al. 2016,  Karpen et al. 2017)  predict that reconnection at the null launches torsional Alfv\'en waves  travelling along the outer spine. The waves are driven by the magnetic twist accumulated elsewhere and released at the reconnection point with the magnetic torque as a restoring force, consistent with the present observation that the dominant oscillatory power moves from a footpoint (C in Fig. 4) to the inner spine (A) at the eruption.

\begin{figure}[tbh]  
\includegraphics[scale=1.0]{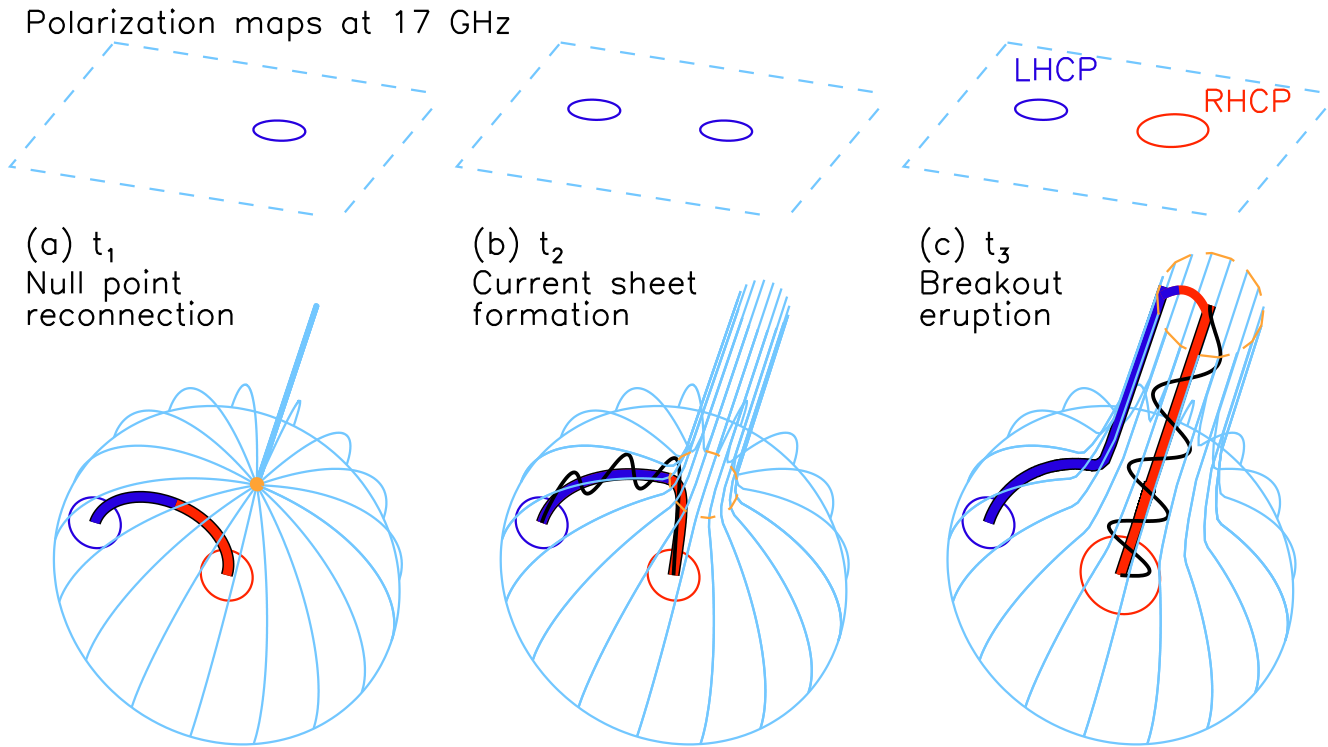}
\caption{Schematic illustration of the observed polarization intensity map (top) and magnetic field configuration (bottom). Thick blue (red) lines stand for a flux rope in the negative (positive) magnetic polarity, sky-blue lines for the fan fields, orange dot for a null point, dashed circles for the breakout current sheet, and the curvy lines for the loop oscillations.}
\end{figure}

\section{Concluding Remarks}

We present three specific phenomena of this microwave CRF: (1) the nonthermal preflare activation, (2) the sudden change of local polarization during the flare, and (3) the oscillation before and during the eruption, as the characteristic features of magnetic reconnection in a fan-spine morphology.
Among these the most obvious evidence for the eruption is the sudden and permanent change of the 17 GHz polarization in the AR center.  The fan-like structure is implied by the 17 GHz preflare emission appearing as a single polarization state over the region with mixed magnetic polarity. The polarization change restricted to the core region then implies that the central part breaks out, letting the inner spine field erupt and revealing a structural change around the inner spine associated with magnetic eruption out of the fan--spine system. This conclusion is solely based on observation and does not refer to a particular model.

The other two pieces are connected to this eruption under the afore-mentioned models designed for a fan-spine reconnection (Karpen et al. 2012, 2017; Pariat et al. 2009, 2010, 2015, 2016;  Wyper et al. 2016, 2017, 2018). 
Especially, we interpreted the post-eruption oscillations in favor of torsional Alfv\'en waves based on specific models (Wyper et al. 2016, Karpen et al. 2017), although briefly discussed other modes as well.
We want to stress here that the most crucial part in this argument is not the exact mode, but the coupling of the preflare and postflare oscillations. The latter judged by their comparable periods implies the transfer of the oscillatory power from the closed loop to the open fields.
The nonthermal activation time, $t_2$, at which the oscillation also starts is another important signature for the breakout current sheet formation, since most of these models predict  transformation of a null point to a current sheet before the eruption (Karpen et al. 2012, 2017;  Wyper et al. 2017, 2018). In this context, the observed time gap $t_3-t_2\approx$ 4 min must corrspond to the time interval between the BCS formation and eruption. Both results are consistent with the models of breakout eruption.  Note that the deformation of the null point to a current sheet is the very essence of the 3D reconnection. On this basis, we argue that these three observed properties are inter-related and evidence for 3D reconnection in the fan--spine structure.

\acknowledgements
JL was supported by NASA grants, 80NSSC18K1705 and 80NSSC18K0673, and NSF grants, AGS 1821294 and AGS 1927578.
SW thanks ISEE/Nagoya University for their support and hospitality during visits to work on Nobeyama data. 
YC and HN was supported by the NNSFC grants 11973031, 11790303 (11790300), and 41774180, and BL, by NNSFC grant 11761141002.
SM is supported by JSPS KAKENHI Grant Number JP18H01253.
Nobeyama Radioheliograph is operated by the International Consortium for the Continued Operation of Nobeyama Radioheliograph (ICCON) consisting of ISEE/Nagoya University, NAOC, KASI, NICT, and GSFC/NASA. 

\appendix
Figure 6 depicts an ideal fan-spine structure evolving from a closed to a partially open structure.  \B{This figure is not included in the paper submitted to the ApJ Letters, but schematically illustrates the proposed scenario.} (a) A perturbation occurs at $t_1$ (Fig, 4b) to cause the null point to deform itself into a current sheet, which is not really violent but gentle to trigger thermal heating mostly in the inner spine (Fig. 4c). The rays emitted from there must have been RHCP, but changes to LHCP while passing through the overlying fan field. (b) The next reconnection occurs at $t_2$  between the magnetic fields inside and outside of the fan across a breakout current sheet, and the microwave source extends to the flux rope. Since this energy release is more intense, the nonthermal phase sets in (Fig. 4c) and the oscillations start everywhere (Fig. 4d). Again, the rays emitted from the two footpoints, initially polarized in different senses, become all LHCP while passing through the fan field. (c) The breakout eruption follows at $t_3$ to cause the flare and let the flux rope erupt. The rays from the positive field will be observed as RHCP as the overlying fan field is now open (Fig. 2). Continuation of the oscillations through this transition (Fig. 4e) implies that they are the magnetic-untwisting waves (torsional Alfven waves) propagating along the spine as guided by the erupted field lines.

\end{document}